\documentclass[conference]{IEEEtran}
\IEEEoverridecommandlockouts
\usepackage{bm}
\usepackage{cite}
\usepackage{amsmath,amssymb,amsfonts}
\usepackage{algorithmic}
\usepackage{graphicx}
\usepackage{textcomp}
\usepackage{xcolor}
\usepackage{subfigure}
\usepackage{float}
\newtheorem{theorem}{Theorem}
\def\BibTeX{{\rm B\kern-.05em{\sc i\kern-.025em b}\kern-.08em
    T\kern-.1667em\lower.7ex\hbox{E}\kern-.125emX}}

\begin{document}

\title{\huge Linear Shrinkage Receiver for Slow Fading Channels under Imperfect Channel State Information}

\author{\IEEEauthorblockN{Wenyi Shi, Shuqin Pang, Wenyi Zhang}
	
\IEEEauthorblockA{Department of Electronic Engineering and Information Science \\
University of Science and Technology of China\\
Emails: \texttt{\{wyshi451, shuqinpa\}@mail.ustc.edu.cn, wenyizha@ustc.edu.cn}}
}

\maketitle

\begin{abstract}
This paper studies receiver design in single-input multiple-output (SIMO) slow fading channels with imperfect channel state information (CSI) at the receiver only. Using generalized mutual information (GMI) as achievable rate, we study the outage behavior when the receiver employs certain generalized form of the nearest neighbor decoding rule. Our study reveals that linearly shrinking the linear minimum mean-squared error (LMMSE) estimate of the CSI reduces the outage probability when the number of receive antennas is finite. Only in the asymptotic regime where the number of receive antennas grows without bound, the LMMSE estimate of the CSI minimizes the outage probability. Numerical results demonstrate that the proposed linear shrinkage receiver achieves evident outage probability reduction.
\end{abstract}

\section{Introduction}
Driven by potential Internet-of-Things (IoT) applications, the fifth-generation (5G) mobile communication systems need to support short-packet communication applications with strict latency and throughput constraints; see, e.g., \cite{shortpackets1}. A slow fading channel model serves as a simple and relevant setup that captures the non-ergodic nature of delay-limited information transmission. A special example is the so-called quasi-static fading model, where the channel fading remains constant during a codeword, but varies from codeword to codeword with respect to a certain probability distribution.

Outage probability is a fundamental performance indicator for such quasi-static fading channels. It characterizes the non-negligible probability that the instantaneous achievable rate is lower than the chosen code rate. The outage behavior of quasi-static fading channels has been thoroughly studied under the ideal assumption of perfect channel state information (CSI) at the receiver. With perfect CSI, the optimal decoder follows a nearest neighbour decoding rule, which selects the (faded) codeword that has the smallest Euclidean distance to the received signal.

In practical systems, however, the CSI is imperfect. It is typically supplied to the receiver via transmitting a prescribed pilot, and the receiver estimates the CSI from its received (faded and noisy) pilot. Nevertheless, owing to its simplicity, the nearest neighbor decoding rule is usually adopted by the decoder even under imperfect CSI, and therefore the decoder is mismatched to the actual channel; see, e.g., \cite{Lapidoth-Fading} \cite{Weingarten-weighted}.

Based on the generalized mutual information (GMI), the generalized outage probability has been defined and studied \cite{Taufiq-CSIR} \cite{Taufiq-CSIT&CSIR}, which characterizes the asymptotic outage behavior of the nearest neighbor decoding rule in the high signal-to-noise ratio (SNR) limit, for multiple-input multiple-output (MIMO) slow fading channels with imperfect CSI.\footnote{Quasi-static fading channels with imperfect CSI is a very realistic issue frequently met in practice, because a coherence block is generally not long enough and many existing standards insert pilots only sparingly within a coherence block. Nevertheless, some oddness exists for such a model: treating the coding block length as an asymptotically large quantity is necessary for information-theoretic analysis, whereas this mathematical assumption would in term imply adequate resource for channel training so as to render the CSI perfect. So this model should be viewed as a compromise between mathematical modeling and engineering practice.} Performance analysis in the finite length regime under imperfect CSI has also been conducted in \cite{Potterr-finite} \cite{Schiessl}.

A recent study reveals that, by generalizing the nearest neighbor decoding rule to incorporate output processing and codeword scaling, the achievable performance in terms of GMI can be substantially improved, for ergodic fading channels with imperfect CSI \cite{Wang-GNNDR} \cite{Pang-GNNDRMIMO}. This observation motivates us to investigate the potential impact of generalized nearest neighbor decoding for slow fading channels, where the performance is measured in terms of outage probability.

In this paper, we conduct a preliminary study along this direction. Instead of solving for the optimal nonlinear output processing and codeword scaling functions as in \cite{Wang-GNNDR}, we turn to a simple linear shrinkage technique: rather than using the linear minimum mean-squared error (LMMSE) estimate of the CSI, we further linearly shrink the LMMSE estimate, when conducting the nearest neighbor decoding. The shrinkage coefficient is selected to minimize the outage probability.

The concept of shrinkage estimator has been used to estimate the covariance matrix in high dimensions. Well-conditioned estimators that combine the sample covariance matrix and certain more stable statistics are referred to as linear shrinkage or weighted estimators \cite{Yuki-Linearshrinkage}. A shrinkage estimator linearly combining the sample covariance matrix and a diagonal (or spherical) matrix may achieve a desirable tradeoff between bias and variance \cite{ledoit-well} \cite{Chen-Shrinkage}. The situation is somewhat similar in our context. For a non-ergodic slow fading channel with imperfect CSI, the achievable rate, i.e., GMI in our study, is in fact a random variable induced by both the true channel fading and the CSI, and its exact value is unknown to the receiver. Although the LMMSE estimator minimizes the mean-squared error of estimated CSI, due to the highly nonlinear relationship between the CSI estimation error and the GMI, the LMMSE estimator does not necessarily achieve the most desirable tradeoff between mean and variance of the GMI, and hence does not necessarily lead to the minimum outage probability. Heuristically, it is risky for the receiver to over-estimate the CSI, because that would lead to an overly optimistic belief about the achievable rate. So a linear shrinkage receiver is potentially helpful in reducing such risk.

Our numerical results suggest that, compared with the LMMSE estimator, the proposed linear shrinkage estimator usually achieves an SNR gain of nearly $1.5$ dB in Rayleigh fading channels with moderate number of receive antennas. On the other hand, in the asymptotic regime where the number of receive antennas grows without bound, we establish that the linear shrinkage estimator degenerates into the LMMSE estimator.

Throughout the paper, the following notation is used. Uppercase bold letters $\mathbf{X}$ represent random vectors and lowercase bold letters $\mathbf{x}$ represent their realizations; uppercase letters $\mathrm{X}$ and lowercase letters $\mathrm{x}$ represent random scalars and their realizations, respectively. Expectation is denoted by $\boldsymbol{E}[\cdot]$. The $n \times n$ identity matrix is $\boldsymbol{I}_{n}$. Symbols $^{*}$ and $\|\|$ represent conjugate transpose and a vector's Euclidean norm, respectively. Logarithm $\log ()$ uses base $e$.

\section{System Model and GMI}

Consider information transmission over a slow fading channel with a single transmit antenna and $N_{r}$ receive antennas, disturbed by independent and identically distributed (i.i.d.) complex additive white Gaussian noise
$\mathbf{Z} \sim \mathcal{CN}\left(\bm{0},\sigma^{2}\boldsymbol{I}_{N_{r}}\right)$. The input-output relationship of the channel is
\begin{equation}
    \mathbf{Y}=\mathbf{S}\mathrm{X}+\mathbf{Z},
    \label{E1}
\end{equation}
where $\mathrm{X} \in \mathbb{C}$ and $\mathbf{S} \in \mathbb{C}^{N_{r} \times 1}$ are the scalar input and the vector channel state (i.e., fading coefficients), respectively. Denote the CSI supplied to the receiver by $\mathbf{V} \in \mathbb{C}^{N_{r} \times 1}$, which is correlated with $\mathbf{S}$. Under the slow fading assumption, $\mathbf{S}$ and $\mathbf{V}$ remain constant over a length-$N$ coding block. Therefore, the conditional probability distribution satisfies
\begin{equation}
    p_{\mathbf{Y} | \mathrm{X}, \mathbf{S}}\left(\mathbf{y}^{N} | \mathrm{x}^{N}, \mathbf{s}\right) = \prod_{n=1}^{N} p_{\mathbf{Y} | \mathrm{X}, \mathbf{S}} \left(\mathbf{y}_{n} | \mathrm{x}_{n}, \mathbf{s}\right).
    \label{E2}
\end{equation}

At the transmitter, when performing coding of a prescribed rate $R$ and block length $N$, the source selects a message $\mathrm{M}$ from $\mathcal{M} = \left\{1,2, \ldots,\left[e^{N R}\right]\right\}$ uniformly randomly and the encoder maps $\mathrm{M}$ to a transmitted codeword $x^N(\mathrm{M})$. We adopt an i.i.d. Gaussian codebook ensemble, for which each codeword obeys $\mathcal{CN}(\bm{0}, P\boldsymbol{I}_N)$, where $P$ is the average transmit power.

At the receiver, let there be (possibly vector-valued) mappings $g$ and $f$ as the output processing function and the codeword scaling function, respectively \cite{Wang-GNNDR}. The decoding metric of the generalized nearest neighbor decoding rule is hence given by\footnote{In \cite{Wang-GNNDR} $g$ and $f$ are scalar-valued, but its derivation of the GMI remains essentially the same for vector-valued $g$ and $f$, which will be adopted in our linear shrinkage estimator.}
\begin{equation}
    D(m)= \frac{1}{N} \sum_{n=1}^{N} \|g\left(\mathbf{y}_{n}, \mathbf{v}\right) - f\left(\mathbf{y}_{n}, \mathbf{v}\right) \mathrm{x}_{n}(m)\|^{2}.
    \label{E3}
\end{equation}

For an i.i.d. codebook ensemble and a prescribed decoding metric, an achievable rate is given by the GMI, which is a lower bound of the mismatch capacity, and is indeed the maximum information rate to guarantee that the decoding error probability, averaged over the specified i.i.d. codebook ensemble, decays to zero as the coding block length grows without bound \cite{Lapidoth-Fading} \cite{Weingarten-weighted}. Below we provide an outlined derivation of the GMI in our problem setup, and for further details, see, e.g., \cite{Wang-GNNDR}.

The decoding error probability averaged over the i.i.d. Gaussian codebook ensemble is identical for all messages due to the symmetry. Therefore, it suffices to study the decoding error probability conditioned upon message $m = 1$ being transmitted. For $m = 1$, we have
\begin{equation}
    \begin{aligned}
    \lim _{N \rightarrow \infty} D(1) &=\lim _{N \rightarrow \infty} \frac{1}{N} \sum_{n=1}^{N}\|g\left(\mathbf{y}_{n}, \mathbf{v}\right) - f\left(\mathbf{y}_{n}, \mathbf{v}\right) \mathrm{x}_{n}(1)\|^{2}\\
    &=\boldsymbol{E}\left[\|g(\mathbf{Y}, \mathbf{V}) - f(\mathbf{Y}, \mathbf{V}) \mathrm{X}\|^{2} \big| \mathbf{S}, \mathbf{V}\right],
    \label{E4}
    \end{aligned}
\end{equation}
almost surely, according to the law of large numbers. Note that here a key difference between the counterpart analysis in \cite{Wang-GNNDR} is that since the channel is non-ergodic, the expectation needs to be conditioned upon $(\mathbf{S}, \mathbf{V})$; in other words, the limit of $D(1)$ is a random variable induced by the channel fading and the CSI encountered.

The GMI is the asymptotic exponent of the probability that an incorrect codeword accumulates a decoding metric smaller than \eqref{E4}, and is obtained from a large deviations analysis, as
\begin{equation}
    I_{\mathrm{GMI}} = \sup_{\theta < 0} \left\{\theta \boldsymbol{E}\left[\|g(\mathbf{Y}, \mathbf{V}) - f(\mathbf{Y}, \mathbf{V}) \mathrm{X}\|^{2} \big| \mathbf{S}, \mathbf{V}\right] - \Lambda(\theta)\right\},
    \label{E5}
\end{equation}
\begin{equation}
    \begin{gathered}
    \Lambda(\theta) =\lim_{N \rightarrow \infty} \frac{1}{N} \Lambda_{N}(N \theta), \\
    \Lambda_{N}(N \theta) =\log \boldsymbol{E}\left[e^{N \theta D(m)} \big| \mathbf{Y}, \mathbf{V}\right], \quad \forall m \neq 1.
    \label{E6}
    \end{gathered}
\end{equation}

Conditioned upon the channel output $\mathbf{Y}$ and the CSI $\mathbf{V}$, $\boldsymbol{E}\left[e^{N \theta D(m)} \mid \mathbf{Y}, \mathbf{V}\right]$ becomes finite product series of the conditional expectations of conditionally independent random variables, wherein $\|g(\mathbf{Y}, \mathbf{V}) - f(\mathbf{Y}, \mathbf{V}) \mathrm{X}\|^{2}$ obeys a non-central chi-squared distribution. We can hence derive that\footnote{Here the expression is slightly more general than that in \cite{Wang-GNNDR} due to the fact that $g$ and $f$ are vector-valued.}
\begin{equation}
    \begin{aligned}
    I_{\mathrm{GMI}}&=\sup_{\theta<0} \{ \theta \boldsymbol{E}\left[\|g(\mathbf{Y}, \mathbf{V}) - f(\mathbf{Y}, \mathbf{V}) \mathrm{X}\|^{2} \big| \mathbf{S}, \mathbf{V}\right] \\
    &-\theta \boldsymbol{E}\left[\|g(\mathbf{Y}, \mathbf{V})\|^{2} + \frac{\theta |g^*(\mathbf{Y}, \mathbf{V}) f(\mathbf{Y}, \mathbf{V})|^{2} P}{1-\theta\|f(\mathbf{Y}, \mathbf{V})\|^{2} P} \Big| \mathbf{S}, \mathbf{V}\right] \\
    &\left.+\boldsymbol{E}\left[\log \left(1-\theta\|f(\mathbf{Y}, \mathbf{V})\|^{2} P\right) \big| \mathbf{S}, \mathbf{V}\right]\right\}.
    \label{E7}
    \end{aligned}
\end{equation}
Again, note that due to non-ergodic fading, the expectations in \eqref{E7} need to be conditioned upon $(\mathbf{S}, \mathbf{V})$, and consequently, the GMI $I_{\mathrm{GMI}}$ is a random variable induced by $\mathbf{S}$ and $\mathbf{V}$. For a code rate $R$, the outage probability can then be written as
\begin{equation}
    P_\mathrm{out} = p\left(I_{\mathrm{GMI}} < R\right).
\end{equation}

It is interesting to emphasize that, since $I_{\mathrm{GMI}}$ depends upon both $\mathbf{S}$ and $\mathbf{V}$, its exact value is unknown even to the receiver, unless $\mathbf{S}$ can be determined by $\mathbf{V}$, i.e., the receiver having perfect CSI.

A natural optimization problem is to solve for the optimal $g$ and $f$ that minimize $P_\mathrm{out}$, i.e., for a given $R$,
\begin{equation}
    \min_{g, f} p\left(I_{\mathrm{GMI}} < R\right).
    \label{E8}
\end{equation}
But this problem appears elusive, partly because the probability distribution of $I_{\mathrm{GMI}}$ is hardly tractable. In the subsequent analysis, we turn to a linear shrinkage estimator of the CSI, which is a special form of $(g, f)$, i.e., $g(\mathbf{Y}, \mathbf{V})$ degenerating into $\mathbf{Y}$ and $f(\mathbf{Y}, \mathbf{V})$ degenerating into ${b}\mathbf{V}$, where ${b}$ is the linear shrinkage coefficient.\footnote{The resulting decoder is different from the weighted nearest neighbor decoder in \cite{Weingarten-weighted}: here only $f$ is scaled, whereas for the weighted nearest neighbor decoder both $g$ and $f$ are simultaneously scaled.}

\section{Linear Shrinkage Receiver}\label{sec:LSR}

In this section, we consider the scenario where the CSI is supplied to the receiver via transmitting a prescribed pilot, as
\begin{equation}
    \mathbf{V} = \mathbf{Y}_{p} = \mathbf{S} X_{p} + \mathbf{Z}_{p},
    \label{E9}
\end{equation} 
where $\mathbf{Y}_{p}$ is the received pilot signal and $X_{p}$ is the transmitted pilot known to both transmitter and receiver. For simplicity, we assume that the covariance matrix of $\mathbf{S}$ is $\eta^{2}
\boldsymbol{I}_{N_{r}}$ and the noise is white Gaussian, $\mathbf{Z}_{p} \sim \mathcal{CN}\left(\bm{0}, \sigma_{p}^{2}
\boldsymbol{I}_{N_{r}}\right)$. According to standard linear estimation theory \cite{Poor1994An}, the LMMSE estimate of $\mathbf{S}$ is given by
\begin{equation}
    \hat{\mathbf{S}} = \frac{\eta^{2} X_{p}^{*} \mathbf{Y}_{p}}{\eta^{2}\left|X_{p}\right|^{2}+\sigma_{p}^{2}}.
    \label{E10}
\end{equation}
If we further assume $\mathbf{S}$ to be complex Gaussian, i.e., Rayleigh fading, then the above LMMSE estimate is also the MMSE estimate $\boldsymbol{E}\left[\mathbf{S} | \mathbf{Y}_{p}\right]$.

Let us denote ${\frac{\eta^{2} X_{p}^{*} }{\eta^{2}\left|X_{p}\right|^{2}+\sigma_{p}^{2}}}$ in the LMMSE estimate by $a$, so that $\hat{\mathbf{S}}_\mathrm{LMMSE} = a \mathbf{Y}_p$. On the other hand, the linear shrinkage estimator with shrinkage coefficient ${b}$ is simply $\hat{\mathbf{S}}_\mathrm{LS} = {b} \mathbf{Y}_p$. Clearly, when ${b} = a$, the linear shrinkage estimator degenerates into the LMMSE estimator. The decoding rule becomes
\begin{equation}
   \hat{m}=\operatorname{argmin}_{m \in \mathcal{M}}\frac{1}{N} \sum_{n=1}^{N}\left\|\mathbf{y}_{n}-{b} \mathbf{v}_{n}\mathrm{x}_{n}(m)\right\|^{2}.
   \label{E11}
\end{equation}

Applying the GMI analysis in the previous section, we obtain the GMI achieved by a receiver employing linear shrinkage estimator.

\begin{theorem}\label{thm:GMI}
    For i.i.d. Gaussian codebook ensemble and generalized nearest neighbor decoding using linear shrinkage estimator, the GMI is given by
    \begin{equation}
        I_{\mathrm{GMI}}=\sup _{\theta<0} K_{\text {LS}},
        \label{E12}
    \end{equation}
    \begin{equation}
        \begin{aligned}
        K_{\text {LS }}&=\theta\left(P\|\mathbf{S}-{b} \mathbf{V}\|^{2}-P\|\mathbf{S}\|^{2}\right)+\log \left(1-P \theta\|{b} \mathbf{V}\|^{2}\right)\\
        &-\frac{P \theta^{2}\left(\|{b} \mathbf{V}\|^{2} \sigma^2 + P\left|\mathbf{S}^{*} {b} \mathbf{V}\right|^{2}\right)}{1-P \theta\|{b} \mathbf{V}\|^{2}}.
        \label{E13}
        \end{aligned}
    \end{equation}
\end{theorem}

{\it Proof:} Specializing $g(\mathbf{y}, \mathbf{v}) = \mathbf{y}$ and $f(\mathbf{y}, \mathbf{v}) = {b} \mathbf{v}$ in \eqref{E7} leads to \eqref{E12} and \eqref{E13}. $\Box$

We remark that $K_\mathrm{LS}$ is a random variable induced by $\mathbf{S}$ and $\mathbf{V}$. Nevertheless, the choice of the parameter $\theta < 0$ can depend upon $\mathbf{S}$ and $\mathbf{V}$. This is because this parameter is from a large deviations analysis and can be any negative real number. The numerator of $\frac{d K_{\text {LS}}}{d \theta}$ is a quadratic function of $\theta$ whose parabola opens upward. Letting $\frac{d K_{\text {LS}}}{d \theta}$ be zero, we can find that either there is a unique negative solution given by the corresponding quadratic formula and is the optimal $\theta$ to maximizing $K_\mathrm{LS}$, or there is no solution and then we should set $\theta \rightarrow 0$ (leading to $I_\mathrm{GMI} = 0$). A special case is when ${b} \mathbf{v} = \mathbf{s}$, i.e., perfect CSI, and then $\theta = -1/\sigma^2$ is the optimal solution, leading to the familiar result of $I_{\mathrm{GMI}} = \log \left(1+P\|\mathbf{S}\|^{2}/\sigma^2\right)$.

Having obtained the optimal $\theta$ for a fixed ${b}$, the linear shrinkage coefficient ${b}$ should be further optimized to minimize the outage probability for each given code rate $R$. That is, the criterion of designing ${b}$ is
\begin{equation}
    \min_{b} p\left(\sup _{\theta<0} K_{\text {LS }} < R\right).
    \label{E14}
\end{equation}
This optimization can be solved numerically via a standard line search procedure; see the discussion in Section \ref{sec:experiment}.

\begin{figure*}[ht]
	\subfigure[$N_r = 4$,  $R = 1$ (bit/channel use)]{
		\begin{minipage}[t]{0.3\linewidth}\label{fig1a}
			\centering
			\includegraphics[width=2.4in]{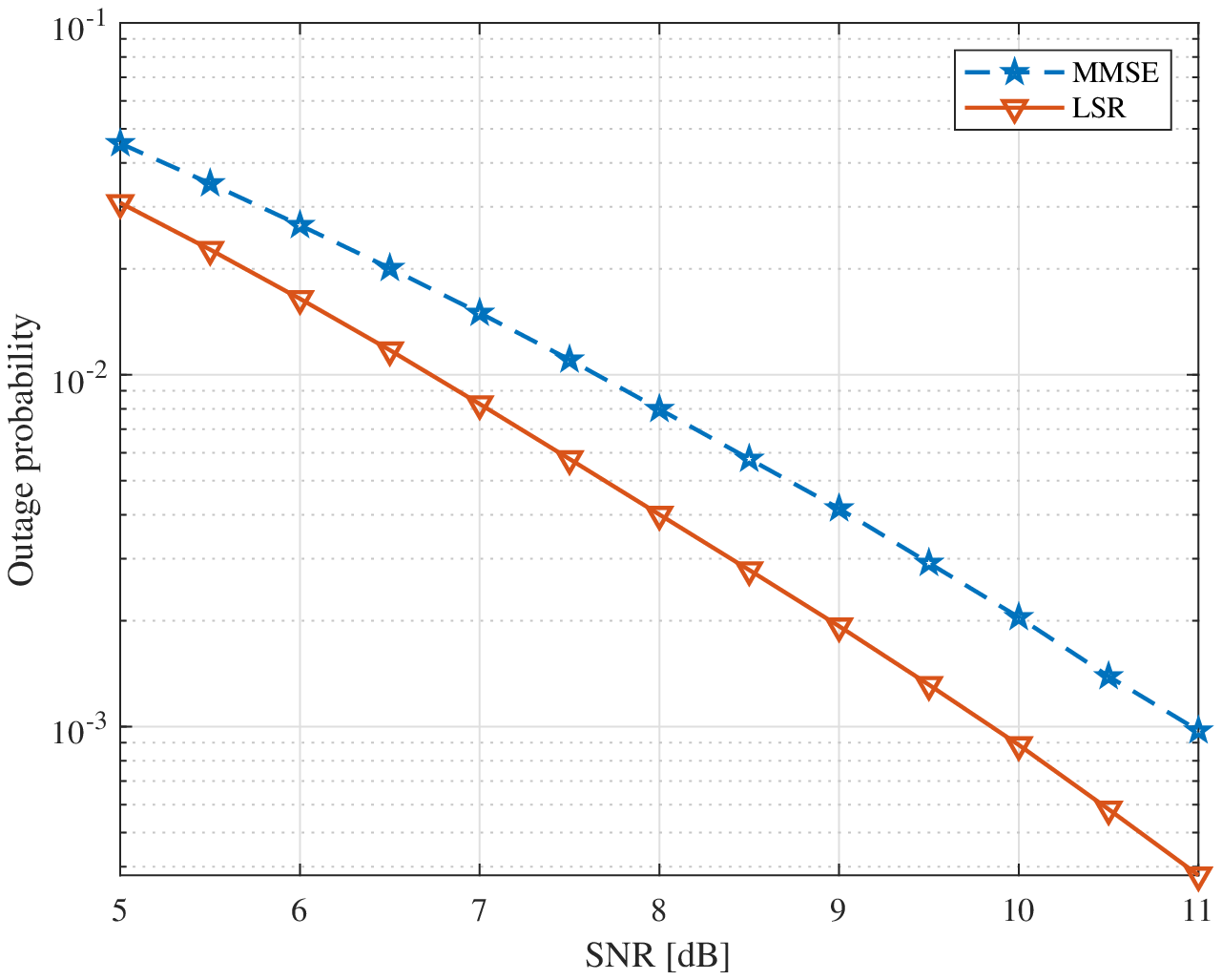}
	\end{minipage}}
	\subfigure[$N_r = 8$,  $R = 2$ (bits/channel use)]{
		\begin{minipage}[t]{0.3\linewidth}\label{fig1b}
			\centering
			\includegraphics[width=2.4in]{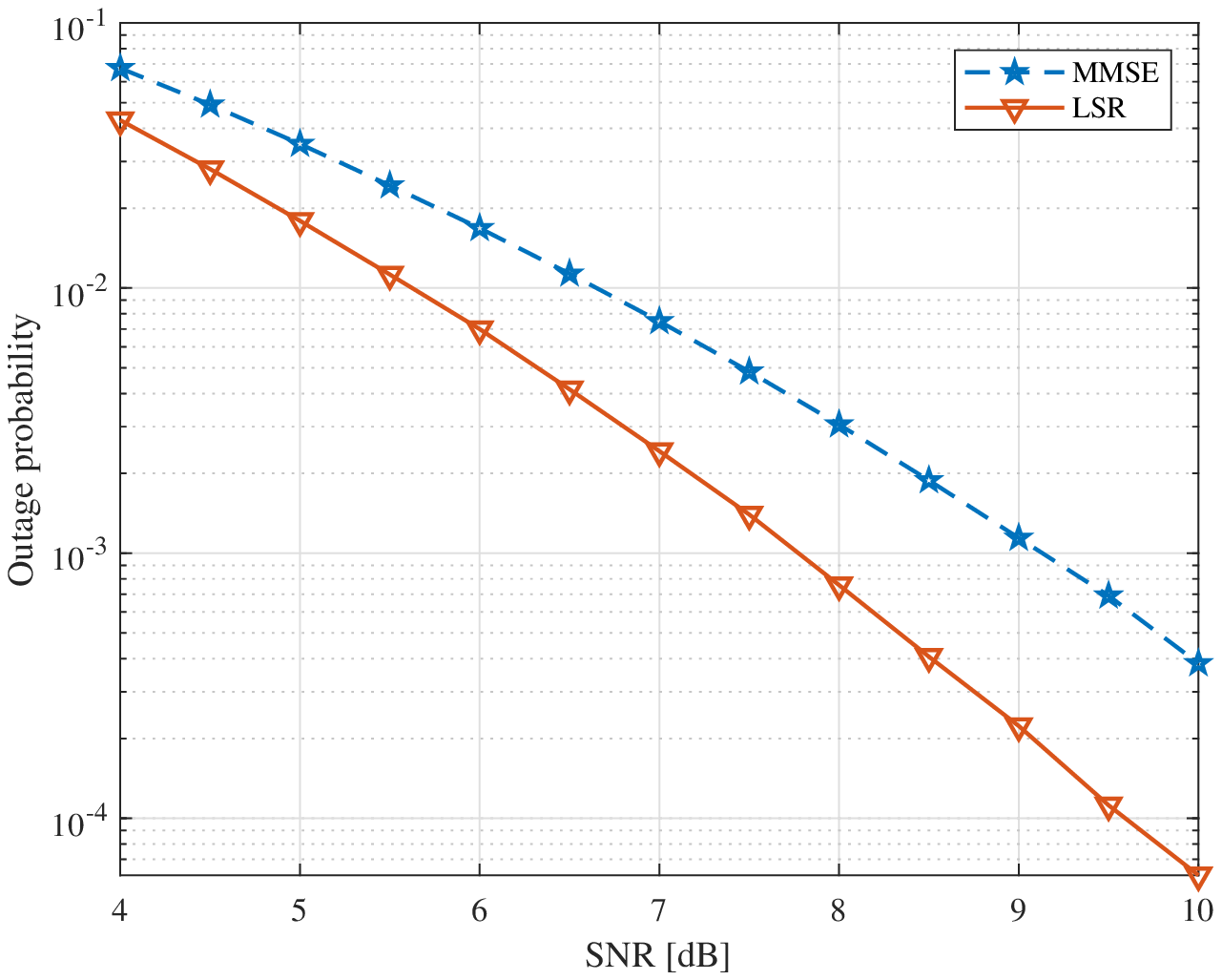}
	\end{minipage}}
	\subfigure[$N_r = 16$,  $R = 3$ (bits/channel use)]{
		\begin{minipage}[t]{0.3\linewidth}\label{fig1c}
			\centering
			\includegraphics[width=2.45in]{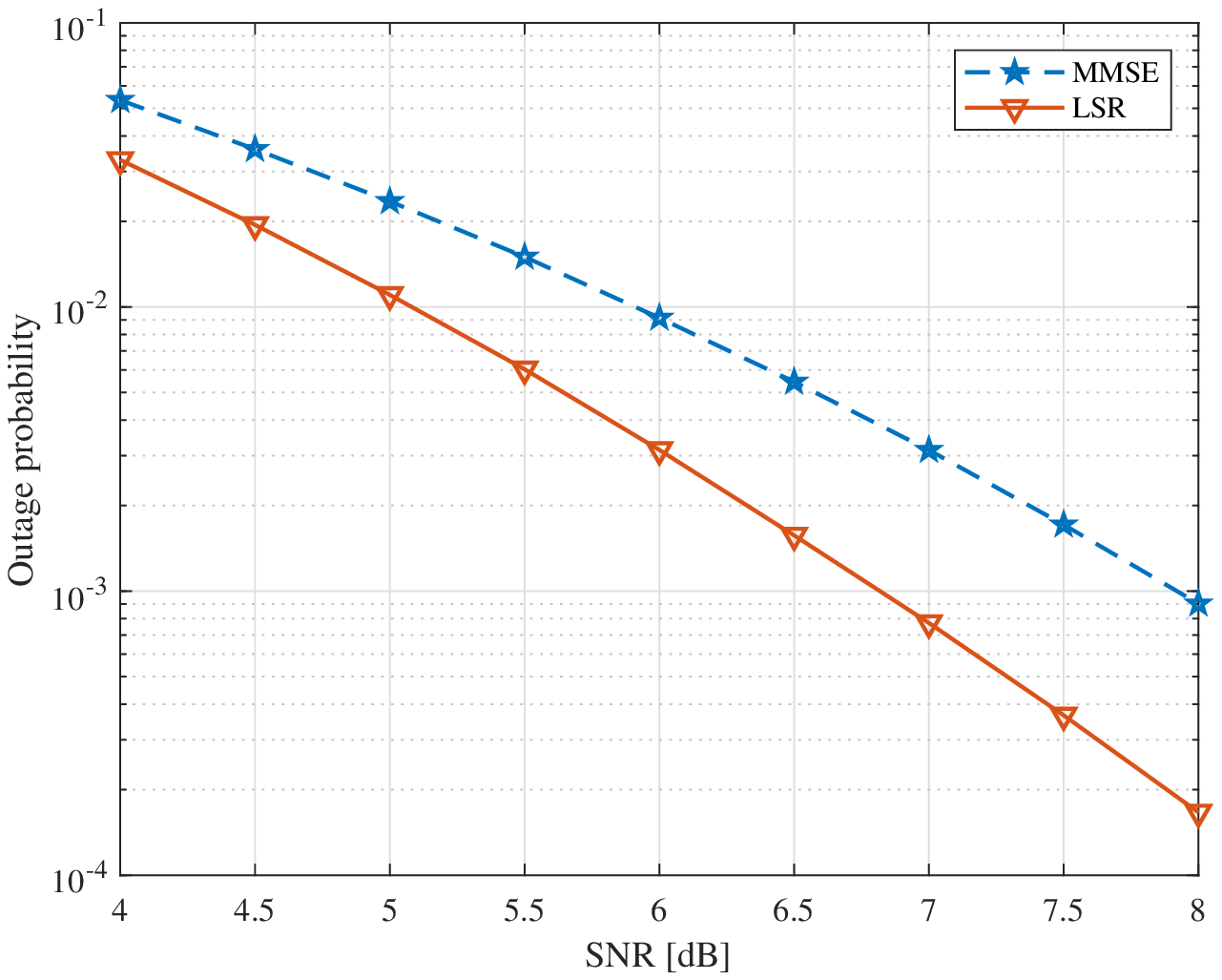}
	\end{minipage}}
	\caption{Outage probability versus SNR under different receive antennas and code rates.}
	\label{fig1}
\end{figure*}

\section{Massive-antenna Asymptotic Analysis}
\label{sec:Massive}

In this section, we investigate how the linear shrinkage estimator behaves asymptotically in the massive-antenna regime. Our main finding is the following result.

\begin{theorem}
    \label{thm:Massive}
    When $b = a$, for any rate $R$, as ${N_{r} \rightarrow \infty}$, the outage probability asymptotically decreases to zero; for any $b \neq a$, there exists a finite threshold rate $\underline{R}$, such that for any rate $R > \underline{R}$, as ${N_{r} \rightarrow \infty}$, the outage probability is bounded away from zero.
\end{theorem}

{\it Proof:} In the massive-antenna regime, we can apply the central limit theorem to obtain the following convergence property: for any $\epsilon > 0$, there exists a corresponding finite $\gamma_\epsilon$, such that for all sufficiently large $N_r$, in
\begin{equation}
    \begin{aligned}
    \|\mathbf{S}\|^{2} = \eta^2 N_r + \eta^2 \sqrt{N_r} \mathrm{W}_\mathbf{s},
    \end{aligned}
\end{equation}
\begin{equation}
    \begin{aligned}
    \|\mathbf{Z}_p\|^{2} = \sigma_p^2 N_r + \sigma_p^2 \sqrt{N_r} \mathrm{W}_\mathbf{z},
    \end{aligned}
\end{equation}
\begin{equation}
    \begin{aligned}
    \mathbf{S}^{*} \mathbf{Z}_{p} = \sqrt{\sigma_p^2 \eta^2 N_r} \mathrm{W}_\mathbf{sz},
    \end{aligned}
\end{equation}
the bounds $- \gamma_\epsilon < \mathrm{W}_\mathbf{s}, \mathrm{W}_\mathbf{z}, \mathrm{W}_\mathbf{sz} < \gamma_\epsilon$ simultaneously hold with probability at least $1 - \epsilon$. Denote the event that the above convergence property holds by $\mathcal{T}$.

For the expression of the GMI in Theorem \ref{thm:GMI}, as mentioned in Section \ref{sec:LSR}, the optimal $\theta$ exists and can be given by a quadratic formula. Its exact form is as follows:\footnote{Throughout the proof we set $\sigma^2 = 1$.}
\begin{equation}
    \begin{aligned}
    \theta^{*} = \frac{-\mathrm{B}-\sqrt{\mathrm{B}^{2}-4 \mathrm{AC}}}{2\mathrm{A}},
    \label{E15}
    \end{aligned}
\end{equation}
where $\mathrm{A}=P^{2} \mathrm{U}_1\|{b}\mathbf{V}\|^{4}+\mathrm{U}_2 P\|{b}\mathbf{V}\|^{2}$, $\mathrm{B}=P\|{b}\mathbf{V}\|^{4}-2 \mathrm{U}_2-2 P \mathrm{U}_1\|{b}\mathbf{V}\|^{2}$, $\mathrm{C}=\mathrm{U}_1-\|{b}\mathbf{V}\|^{2}$,
$\mathrm{U}_1=\|\mathbf{S}-{b}\mathbf{V}\|^{2}-\|\mathbf{S}\|^{2}$, and $\mathrm{U}_2=\|{b}\mathbf{V}\|^{2}+P\left|\mathbf{S}^{*} {b}\mathbf{V}\right|^{2}$. Utilizing the expression of $\mathbf{V}$ in terms of $\mathbf{S}$ and $\mathbf{Z}_{p}$, we have
\begin{equation}
    \begin{aligned}
    \|\mathbf{V}\|^{2}=\|\mathbf{S}\|^{2}\left|X_{p}\right|^{2}+\left\|\mathbf{Z}_{p}\right\|^{2}+\mathbf{S}^{*}X_{p}^{*} \mathbf{Z}_{p}+\mathbf{Z}_{p}^{*} \mathbf{S}X_{p},
    \end{aligned}
\end{equation}
\begin{equation}
    \begin{aligned}
    \mathbf{S}^{*} {b} \mathbf{V}+{b}^{*} \mathbf{V}^{*} \mathbf{S}=\|\mathbf{S}\|^{2}\left({b}X_{p}+{b}^{*} X_{p}^{*}\right)+{b}\mathbf{S}^{*} \mathbf{Z}_{p}+{b}^{*} \mathbf{Z}_{p}^{*} \mathbf{S}.
    \end{aligned}
\end{equation}
%\begin{equation}
%    \begin{aligned}
%    \left|\mathbf{S}^{*} \mathbf{V}\right|^{2}=\|\mathbf{S}\|^{4}\left|X_{p}\right|^{2}+\left|\mathbf{S}^{*} \mathbf{Z}_{p}\right|^{2}
%    +\|\mathbf{S}\|^{2}\left(\mathbf{S}^{*}X_{p}^{*} \mathbf{Z}_{p}+\mathbf{Z}_{p}^{*} \mathbf{S}X_{p}\right).
%    \end{aligned}
%\end{equation}
That is, $\theta^*$ can be represented in terms of $\|\mathbf{S}\|^2$, $\|\mathbf{Z}_p\|^2$, and $\mathbf{S}^* \mathbf{Z}_p$.

Now, we let ${b}=a=\frac{\eta^{2}X_{p}^{*}}{\eta^{2}\left|X_{p}\right|^{2}+\sigma_{p}^{2}}$, and proceed to analyze the asymptotic behavior of the resulting $\theta^*$, conditioned on event $\mathcal{T}$.

A long and tedious calculation shows that, when ${b} = a$, the $N_r^3$ and $N_r^{2.5}$ terms in the denominator $2\mathrm{A}$ of $\theta^*$ vanish, and the remaining positive highest order term is $O(N_r^2)$. On the other hand, the numerator of $\theta^*$ is dominated by $-2\mathrm{B}$, and its leading term is also $O(N_r^2)$. Consequently, when ${b} = a$, $\theta^*$ is strictly negative, and the dominant term in the corresponding $K_\mathrm{LS}$, i.e., \eqref{E13}, is given by
\begin{equation}
    \begin{aligned}
    K_{\text {LS}} = \log \left(1-P \theta^{*} \frac{\eta^{4}\left|X_{p}\right|^{2}}{\eta^{2}\left|X_{p}\right|^{2}+\sigma_{p}^{2}} N_{r}\right),
    \end{aligned}
\end{equation}
which is $O\left(\log{N_{r}}\right)$.

Then, consider any ${b} \neq a$, and investigate the asymptotic behavior of $\theta^*$, also conditioned on $\mathcal{T}$. Now the denominator $2\mathrm{A}$ of $\theta^*$ has a nonzero leading term of order $N_r^3$, and consequently, $\theta^* = O(1/N_r)$. Evaluating $K_\mathrm{LS}$ in \eqref{E13}, we can show that there exists a constant $\overline{K}$ independent of $N_r$ upper bounding $K_\mathrm{LS}$, i.e., $K_\mathrm{LS} \leq \overline{K} = O(1)$.

So, in order to conclude the proof, we have for any given $R$ and $\epsilon > 0$:

(a) If ${b} = a$,
\begin{equation}
    \begin{aligned}
        p(\sup_{\theta < 0} K_\mathrm{LS} < R) &= p(\sup_{\theta < 0} K_\mathrm{LS} < R | \mathcal{T}) p(\mathcal{T})\\
        &+ p(\sup_{\theta < 0} K_\mathrm{LS} < R | \mathcal{T}^c) p(\mathcal{T}^c)\\
        &\leq p(O(\log N_r) < R) + p(\mathcal{T}^c) \rightarrow 0
    \end{aligned}
\end{equation}
as $N_r \rightarrow \infty$.

(b) If ${b} \neq a$,
\begin{equation}
    \begin{aligned}
        p(\sup_{\theta < 0} K_\mathrm{LS} < R) &= p(\sup_{\theta < 0} K_\mathrm{LS} < R | \mathcal{T}) p(\mathcal{T})\\
        &+ p(\sup_{\theta < 0} K_\mathrm{LS} < R | \mathcal{T}^c) p(\mathcal{T}^c)\\
        &\geq (1 - \epsilon) p(\overline{K} < R),
    \end{aligned}
\end{equation}
as $N_r \rightarrow \infty$, which is bounded away from zero for all rates $R > \overline{K}$. This completes the proof. $\Box$

Theorem \ref{thm:Massive} holds in the massive-antenna limit.\footnote{Theorem \ref{thm:Massive} may appear natural from a heuristic view, but we feel that its underlying mechanism is different from the well known channel hardening effect. Because the CSI is noisy over each individual antenna, even with an infinite number of antennas it is still impossible to decrease the channel estimation error to zero. So according to our GMI analysis, it is not obvious to assert that the LMMSE coefficient is the outage minimizer.} But for a finite number of receive antennas, the optimal shrinkage coefficient ${b}$ may noticeably deviate from $a$, and its reduction in outage probability may be evident, as our numerical experiments suggest, in the next section.

\begin{figure}[ht]
    \centerline{\includegraphics[scale=0.5]{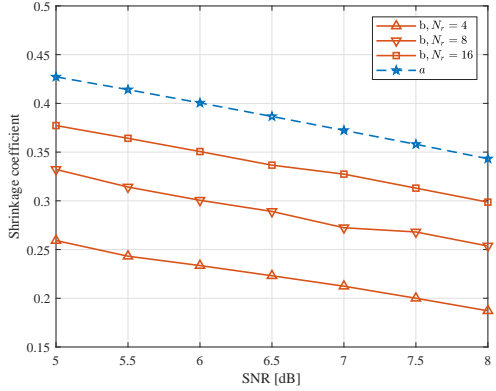}}
    \caption{Shrinkage coefficients versus SNR, under different values of $N_r$. Note that the LMMSE estimate coefficient $a$ does not depend upon $N_r$.}
    \label{fig2a}
\end{figure}

\begin{figure}[ht]
    \centerline{\includegraphics[scale=0.5]{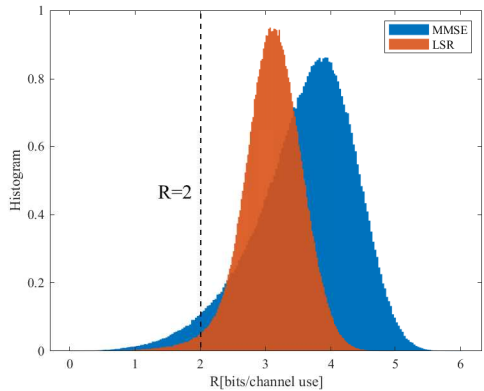}}
    \caption{Comparison of histograms of GMI, under $N_r = 8$, $R = 2$ (bits/channel use), $\mathrm{SNR} = 5$dB.}
    \label{fig2b}
\end{figure}

\begin{figure}[ht]
    \centerline{\includegraphics[scale=0.5]{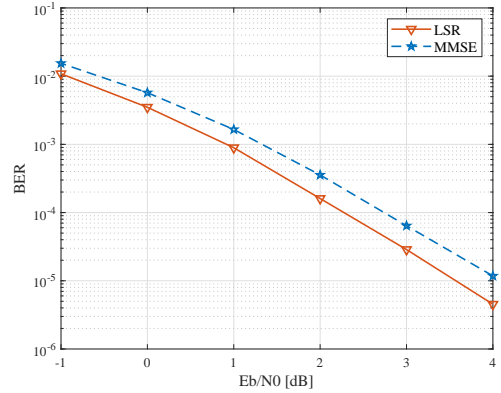}}
    \caption{BER comparison on 5G NR LDPC, under $N_{r} = 16$, $N = 544$, and $R = 1/3$ (bits/channel use).}
    \label{fig3}
\end{figure}

\section{Numerical Simulations}
\label{sec:experiment}

In this section, numerical simulations are conducted to validate the effectiveness of our proposed linear shrinkage receiver. We begin by comparing the outage probability of the LMMSE estimator and the linear shrinkage estimator, and then demonstrate the bit-error-rate (BER) decoding performance on the 5G NR LDPC. Spatially independent Rayleigh fading is adopted throughout, for simplicity.

\subsection{Outage Probability}

For each experiment, we fix the code rate and vary the SNR. We assume that the pilot noise variance $\sigma^2_p$ is identical to the noise variance during information transmission $\sigma^2$, and that the pilot is $X_p = \sqrt{P}$. The LMMSE estimate coefficient $a$ is hence real positive, and we further restrict our search of optimal ${b}$ to be along the real positive line as well.\footnote{This is a simplification based upon our numerical experiment, which shows that allowing ${b}$ to have imaginary part does not appear rewarding.}

Figure \ref{fig1} shows the outage probabilities with SNR, under three settings. Therein, ``LSR'' denotes the performance achieved by a receiver using linear shrinkage estimator. Compared with the LMMSE estimator, the linear shrinkage estimator usually has an SNR gain of nearly $1.5$ dB.

The optimal shrinkage coefficients ${{b}}$ are depicted in Figure \ref{fig2a} for the three settings in Figure \ref{fig1}, and we also plot the LMMSE coefficient $a$. The trend of ${{b}}$ is roughly consistent with that of ${a}$ with SNR. For a fixed SNR, the value of ${{b}}$ is closer to ${a}$ as the number of receive antennas increases. This is in line with our theoretic analysis in Section \ref{sec:Massive}. Figure \ref{fig2b} compares the histograms of the GMI between the linear shrinkage estimator and the LMMSE estimator, under $N_r = 8$, $R = 2$ bits/channel use, and $\mathrm{SNR} = 5$ dB. We observe that the linear shrinkage estimator achieves a more desirable mean-variance tradeoff compared to the LMMSE estimator: the linear shrinkage estimator induces a GMI distribution with a slightly decreased mean but a ``thinned'' tail in the low rate region, so the overall effect is a smaller outage probability.

\subsection{Decoding Performance on 5G NR LDPC}

We evaluate the performance of the linear shrinkage estimator for 5G New Radio low-density-parity-check (LDPC) short codes \cite{Richardson-LDPCNR}. We use QPSK modulation and transmit coded symbols over a slow fading channel. The received signal is decoded via the belief propagation algorithm based on soft decision. The linear shrinkage coefficient affects the log-likelihood ratio (LLR). Figure \ref{fig3} shows the BER performance for a code rate $1/3$ bits/channel use, codeword length $544$, with $16$ receive antennas. The results demonstrate that the linear shrinkage technique is capable of improving the reliability for short codes.

\section{Conclusion}

In this paper, we have proposed to apply a linear shrinkage technique to improve the outage behavior for slow fading channels with imperfect CSI. By linearly shrinking the LMMSE estimate of the CSI, the probability distribution of the GMI can be tuned to yield an improved mean-variance tradeoff, and consequently the outage probability is reduced. This effect is evident when the number of receive antennas is not too large. On the other hand, in the asymptotic regime of massive antennas, the LMMSE estimator is the only linear shrinkage estimator that achieves asymptotically vanishing outage probability at all rates. For future research, an interesting direction is to investigate whether some form of nonlinear processing, beyond the linear shrinkage receiver here, can lead to further performance improvement.

\section*{Acknowledgment}
This work was supported by the National Key Research and Development Program of China under Grant 2018YFA0701603.

\bibliographystyle{IEEEtran}
%\bibliography{ref}

\end{document}